\documentclass[10pt]{iopart}
\newif\ifpdf\ifx\pdfoutput\undefined\pdffalse\else\pdfoutput=1\pdftrue\fi
       \newcommand{\pdfgraphics}{\ifpdf\DeclareGraphicsExtensions{.pdf,.jpg}\else\fi}
\usepackage{graphicx}
\usepackage{hyperref}
\usepackage{amssymb}
\usepackage{longtable}
\usepackage{rotating}
\usepackage{color}
\usepackage{epsfig}
\usepackage{epsf}
\usepackage{fancyhdr}
\def\be{\begin{equation}}
\def\ee{\end{equation}}
\def\th{\textrm{\mbox{\tiny{th}}}}
\def\Tobs{T_{\textrm{\mbox{\tiny{obs}}}}}
\def\Tcoh{T_{\textrm{\mbox{\tiny{coh}}}}}

\def\min{\textrm{\mbox{\tiny{min}}}}
\def\max{\textrm{\mbox{\tiny{max}}}}

\begin{document}
\pdfgraphics
%
\title[Wide parameter search for isolated pulsars using the Hough transform]
{Wide parameter search for isolated pulsars using the Hough transform}
\author{Badri Krishnan\\ (for the LIGO Scientific Collaboration)}
%
\address{\dag\ Max-Planck-Institut f\"ur
    Gravitationsphysik, Albert Einstein Institut, Am M\"uhlenberg 1,
    D-14476 Golm, Germany}


\address{E-mail: \texttt{badri.krishnan@aei.mpg.de} }

\begin{abstract}
  We use the Hough transform to analyze data from the second science
  run of the LIGO interferometers, to look for gravitational waves
  from isolated pulsars. We search over the whole sky and over a large
  range of frequencies and spin-down parameters. Our search method is
  based on the Hough transform, which is a semi-coherent,
  computationally efficient, and robust pattern recognition technique.
  We also present a validation of the search pipeline using hardware
  signal injections.
\end{abstract}


%
%

\section{Introduction}
\label{sec:intro}

This paper presents some partial results for a wide parameter
space search for periodic gravitational waves using data from the
LIGO detectors.  The most promising sources for such waves are
isolated pulsars. Previous searches for gravitational waves from
pulsars have been of two kinds.  The first is a search targeting
pulsars whose parameters are known through radio observations.
These searches typically use matched filtering techniques and are
not very computationally expensive.  An example of such a search
is \cite{S1-CW} which targets PSR J1939+2134 using data from the
first science runs of the LIGO and GEO detectors.  The end result
is an upper limit on the strength of the gravitational wave
emitted by this pulsar and therefore on its ellipticity.  See also
\cite{cw-prl} which applies some of the techniques presented in
\cite{S1-CW} to a large number of known pulsars using data from
the second science run of the LIGO detectors.  The second kind of
searches look for pulsars which have not yet been observed by
radio telescopes. This involves searching over large
parameter space volumes and turns out to be computationally limited.
This is 
because looking for weak continuous wave signals requires large
observation times to build up signal to noise ratio and to claim a
detection with some degree of confidence; on the other hand, the
number of templates that must be searched over, and therefore the
computational requirements, increase rapidly with the observation
time.  An example of such a search is \cite{astone} where a 2-day
long data stretch from the Explorer bar detector is used to
perform an all sky search in a narrow frequency band around the
resonant frequency of the detector.

All the searches mentioned above rely on a coherent integration
over the full observation time; it is well known that this is the
optimal method. However, a full coherent integration is
computationally expensive and it is therefore also useful to
consider methods which are less sensitive but computationally
inexpensive. Such methods typically involve semi-coherent
combinations of the signal power in short stretches of data.
The Hough transform is an example of such a method \cite{hough04}.
Using this method, we perform an {\it all-sky} search over a large
frequency range using two months of data from the LIGO detectors.  As
in all the searches mentioned above, we assume that the pulsar does
not glitch during the full observation time considered.  

Section \ref{sec:waveform} briefly describes the waveforms that we
are looking for.  Section \ref{sec:houghbasics} describes our
search method, the Hough transform. The search pipeline and the
parameter space we search over is given in section
\ref{sec:parameters}.  The search results are given in section
\ref{sec:searchresults}.  Section \ref{sec:hardwareinj} presents a
validation of our search method using hardware injected signals
and finally section \ref{sec:conc} concludes with a summary of our
results and plans for further work.

\section{The expected waveform}
\label{sec:waveform}

The form of the gravitational wave emitted by an isolated pulsar, as seen by a
gravitational wave detector is \cite{jks}:
\begin{equation}
h(t) = F_+(t,\psi)h_+(t) + F_\times(t,\psi)h_\times(t)
\end{equation}
where $t$ is time in the detector frame, $\psi$ is the polarization
angle of the wave, $F_{+,\times}$ are the detector antenna pattern
functions for the two polarizations.  If
we assume the emission mechanism is due to deviations of the pulsar's
shape from perfect axial symmetry, then the gravitational waves are
emitted at a frequency which is twice the rotational rate $f_r$ of the
pulsar.  Under this assumption, the waveforms for the two
polarizations $h_{+,\times}$ are given by:
\begin{equation}
h_+ = h_0 \frac{1+ \cos^2\iota}{2}\cos\Phi(t)\,, \qquad
h_\times = h_0 \cos\iota\sin\Phi(t)\,,
\end{equation}
where $\iota$ is the angle between the pulsar's spin axis and the
direction of propagation of the waves, and $h_0$ is the amplitude:
\begin{equation} \label{eq:h0} h_0 = \frac{16\pi^2G}{c^4}\frac{I_{zz}\epsilon
f_r^2}{d}\,. \end{equation}
Here $d$ is the distance of the star from Earth, $I_{zz}$ is the
$z$-$z$ component of the star's moment of inertia with the $z$-axis being
its spin axis, and $\epsilon$ is the equatorial ellipticity of the
star. The phase $\Phi(t)$ takes its simplest form in the Solar System Barycenter
(SSB) frame where it can be expanded in a Taylor series.  Up to second order:
\begin{equation} \label{eq:phasemodel}
\Phi(t) = \Phi_0 + 2\pi\left( f_0(T - T_0) +
\frac{1}{2}\dot{f}(T-T_0)^2 \right)\,.
\end{equation}
Here $T$ is time in the SSB frame and $T_0$ is a fiducial start time.
The frequency $f_0$ and the spin-down parameter $\dot{f}$ are defined
at this fiducial start time. In this paper, we
include only one spin-down parameter in our search, i.e. we ignore the
higher order terms in Eq.~(\ref{eq:phasemodel}).  This is reasonable
because, as we shall see below in section \ref{sec:parameters}, our
frequency resolution is too coarse for the higher spindowns to have
any effect for reasonable values of the pulsar spindown age.   

Neglecting relativistic effects which do not affect us significantly
in this case (again because of the coarseness of our frequency
resolution), the instantaneous frequency $f(t)$ of the wave as
observed by the detector is given, to a very good approximation, by
the familiar non-relativistic Doppler formula:
\begin{equation}\label{eq:master}
f(t) - \hat{f}(t) = \hat{f}(t)\frac{ {\bf v} (t)\cdot\bf{n}}{c}
\end{equation}
where $t$ is time in the detector frame, ${\bf v}(t)$ is the
velocity of the detector at time $t$, $\bf{n}$ is the direction to
the pulsar, $\hat{f}(t)$ is the instantaneous signal frequency at
time $t$ and is given by
\begin{equation} \label{eq:fhat}
\hat{f}(t) = f_0 + \dot{f}\left(t -t_0 \right)\,.
\end{equation}
Equations (\ref{eq:master}) and (\ref{eq:fhat}) describe the
time-frequency pattern produced by a signal, and this is the pattern
that the Hough transform is used to look for.

\section{The Hough transform}
\label{sec:houghbasics}

The Hough
transform was invented by Paul Hough in 1959 as a method for finding
patterns in bubble chamber pictures from CERN \cite{hough1} and it was
later patented by IBM \cite{hough2}.  The Hough
transform is also well known in the literature on pattern recognition to be
a robust method for detecting straight lines, circles etc. in digital
images; see e.g. \cite{ik} for a review in this field.  A detailed discussion
of the Hough transform as applied to the search for continuous
gravitational waves can be found in \cite{hough04}.  A closely related
semi-coherent method is the stack-slide algorithm described in
\cite{bc}.

The idea of the Hough transform can be illustrated by the following
simple example.  Consider the problem of trying to detect straight
lines in a noisy two-dimensional digital image.  The digital image is
assumed to be made up of pixels which can be in one of only two
possible states, namely ``on'' or ``off''.  Let $(x,y)$ be the coordinates
of the center of a typical pixel.  We are looking for a pattern which
is parameterized by two numbers $(m,c)$ such that
\begin{equation}
y = mx + c\,.
\end{equation}
The parameter space $(m,c)$ is assumed to be suitably digitized so
that it is also made up of pixels.
To find the most likely value of $(m,c)$, we proceed as follows.  For
each pixel $(\hat{x},\hat{y})$ which is ``on'', we mark all the
possible values of $(m,c)$ which are consistent with it, i.e. we mark
all pixels in the $(m,c)$ plane lying on the straight line $\hat{y} =
m\hat{x} + c$ with a ``+1''.  This is repeated for every pixel which is
``on''.   The end result is an integer, the number count, for every
pixel in the $(m,c)$ plane.  In the case when the digital image is too
noisy and no straight lines can be detected, the number counts would be
uniformly distributed in the $(m,c)$ place.  The presence of a
sufficiently strong signal would lead to a large number count in at
least one of the $(m,c)$ pixels, and the largest number count would
indicate the most likely parameter space values.

This method enables us to mark all the possible templates consistent
with a given observation without stepping through the parameter space
point-by-point. This leads to a significant gain in computational
speed. Furthermore, each observation, no matter how noisy, only adds
at the most $+1$ to the final number count.  These two features are
the chief virtues of the Hough transform method: computational speed
and robustness.  On the other hand, the Hough search is likely to be
less sensitive than the stack-slide search considered in \cite{bc}.
The tradeoffs between sensitivity versus efficiency and robustness
are yet to be studied in detail and will be important in the context
of a hierarchical search \cite{bc, cgk}.  

In our case, the Hough transform is used to find a
signal whose frequency evolution fits the pattern produced by the
Doppler shift (\ref{eq:master}) and the spin-down (\ref{eq:fhat}) in
the time-frequency plane.  The parameters which determine this pattern
are $(f_0,\dot{f},\mathbf{n})$; a point in this four dimensional
parameter space will be denoted by $\vec{\xi}$.  This parameter space is
covered by a discrete cubic grid whose resolution is described in
section \ref{sec:parameters}.  The result of the Hough transform is a
histogram, i.e. an integer (the
number count) for each point of this grid.  The starting point for the
Hough transform are $N$ short stretches of Fourier transformed data; each
short stretch will be called an SFT (Short Fourier Transform).  Each
of these SFTs is ``digitized'' by setting a threshold $\rho_\th$ on
the normalized power $\rho_k$ in the $k^{th}$ frequency bin:.
\begin{equation} \label{eq:normpower}
\rho_k = \frac{2|\tilde{x}_k|^2}{\Tcoh S_n(f_k)} \,.
\end{equation}
Here $\tilde{x}_k$ is the value of the Fourier transform in the
$k^{th}$ frequency bin corresponding to a frequency $f_k$, $\Tcoh$ is
the time baseline of the SFT, and $S_n(f_k)$ is the single-sided power
spectral density of the detector noise at the frequency $f_k$.  We
require that $\Tcoh$ is small enough so that the signal does not shift
by more than, say, half a frequency bin within this time duration.
For frequencies of $\sim 300$Hz, this restricts $\Tcoh$ to be lesser
than $\sim 60$min \cite{hough04}.  In this paper, we work with SFTs
for which $\Tcoh = 1800$s.  In principle, we could choose $\Tcoh$ to
be greater, but we are restricted by the duty cycle of the
interferometers in that we should be able to find suitably long
time periods during which the detector is in lock.  Furthermore, the
data should be stationary over the chosen time period.  The choice of
$1800$s is a suitable compromise for all the three interferometers
during the S2 run.    

This thresholding produces a set of zeros and ones (called a
``peakgram'') from each SFT.  This set of peakgrams is the analog of
the digitized two-dimensional image described earlier.  The Hough
transform is used to calculate the number count $n$ at each parameter
space point starting from this collection of peakgrams.  Let $p(n)$ be the
probability distribution of $n$ in the absence of a signal, and
$p(n|h)$ the distribution in the presence of a signal $h(t)$.  It is
clear that $0\leq n \leq N$, where $N$ is the number of SFTs, and it
can be shown that for stationary
Gaussian noise, $p(n)$ is a binomial distribution with mean
$Nq$ where $q = e^{-\rho_\th}$ is the probability that any frequency
bin is selected:
\begin{equation}
\label{eq:binomialnosig}
p(n) = \left( \begin{array}{c} N \\ n \end{array}  \right)
q^n(1-q)^{N-n}\,.
\end{equation}
In the presence of a signal, the distribution is
ideally also a binomial but with a slightly larger mean $N\eta$
where, for weak signals, $\eta$ is given by
\begin{equation}
\eta = q\left\{1+\frac{\rho_\th}{2}\lambda +
\mathcal{O}(\lambda^2)  \right\}\,.
\end{equation}
$\lambda$ is the signal to noise ratio within a single SFT, and
for the case when there is no mismatch between the signal and the
template:
\begin{equation}
\lambda = \frac{4|\tilde{h}(f_k)|^2}{\Tcoh S_n(f_k)}
\end{equation}
with $\tilde{h}(f)$ being the Fourier transform of the signal $h(t)$
(see \cite{hough04} for details). The approximation that the
distribution in the 
presence of a signal is binomial breaks down for reasonably strong
signals.  This happens mainly due to two reasons: i) the random
mismatch between the signal and the template used to calculate the
number count and ii) the amplitude modulation of the signal which
causes $\eta$ to vary from one SFT to another and for different sky
locations. The result of these two effects is to ``smear'' out the
binomial distribution in the presence of a signal.

Candidates in parameter space are selected by setting a threshold
$n_\th$ on the number count.  The false alarm and
false dismissal rates for this threshold are defined respectively in
the usual way:
\begin{equation}
\alpha = \sum_{n=n_\th}^{N} p(n) \,,\qquad
\beta = \sum_{n=0}^{n_\th-1}p(n| h)\,.
\end{equation}
We choose the thresholds $(n_\th,\rho_\th)$ based on the
Neyman-Pearson criterion of minimizing $\beta$ for a given value of
$\alpha$.  It can be shown \cite{hough04} that this criterion leads,
in the case of weak signals (i.e. $\lambda << 1$), large $N$, and
Gaussian stationary noise, to 
$\rho_\th \approx 1.6$.  This corresponds $q = e^{-\rho_\th} \approx
0.20$, i.e. we select about $20\%$ of the frequency bins from each
SFT; for weak signals, this turns out to be independent of the choice
of $\alpha$ and signal strength. Furthermore, $n_\th$ is also
independent of the signal strength and is given by: 
\begin{equation}
n_\th= Nq +
\sqrt{2N q(1-q)}\,\textrm{erfc}^{-1}(2\alpha)
\end{equation}
where, as before, $q=e^{-\rho_\th}$ and $\textrm{erfc}^{-1}$ is the
inverse of the complementary error
function.  These values of the thresholds lead to a certain value of
the false dismissal rate $\beta$ which is given in \cite{hough04}.
The value of $\beta$ of course depends on the signal strength, and
it turns out that the weakest signal which will cross the above
thresholds at a false alarm rate of $1\%$ and a false dismissal
rate of $10\%$ is given by
\begin{equation} \label{eq:sensitivity}
h_0 = \frac{8.54}{N^{1/4}}\sqrt\frac{S_n(f_0)}{\Tcoh}
\end{equation}
Equation (\ref{eq:sensitivity}) gives the smallest signal which can be
detected by the search, and is therefore a measure of the sensitivity
of the search.

The data analyzed in this paper correspond to LIGO's second
Science Run (S2) that was held for 59 days, from February 14 to
April 14, 2003. The GEO detector was not running at that time, but
all three LIGO detectors were operating  with a significantly
better sensitivity than during the first science run. The  LIGO
detectors comprise one 4 km facility in Livingston, Louisiana,
(L1) and two, 4 km and 2 km respectively in Hanford, Washington
(H1 and H2); see eg. \cite{ligo-geo}. For our purposes, we note
that the duty cycles of the detectors during the S2 run were $37\%$
for L1, $74\%$ for H1 and $58\%$ for H2.  The number $N$ of $30$min
SFTs available for L1 data were 687, 1761 for H1 and 1384 for H2. 

\begin{figure}
  \begin{center}
  \includegraphics[height=7cm]{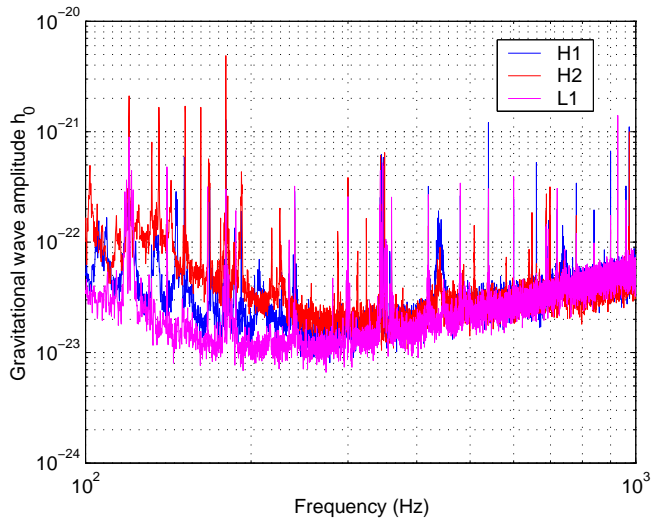}
  \caption{Typical sensitivities of the three LIGO detectors during
  the S2 run with a 1\% false alarm rate and 10\% false dismissal rate.}
  \label{fig:S2sensitivity}
  \end{center}
\end{figure}
Figure \ref{fig:S2sensitivity} shows the expected sensitivity for
the Hough search by the three LIGO interferometers during the S2
run. Those $h_0$ values correspond to the amplitudes detectable
from a generic source with a 1\% false alarm rate and 10\% false
dismissal rate, as given by Eq.~(\ref{eq:sensitivity}).  It should be kept
in mind that Eq.~(\ref{eq:sensitivity}) significantly over-estimates the
sensitivity of the search for unknown pulsars because it does not
include the mismatch between the signal and the template.
Furthermore, due to the large number of templates involved in the
search, a false alarm rate of $1\%$ is too large in practice, and
it would result in too many potential candidates.  A false alarm rate
of $\sim 10^{-13}$ would be more realistic since this would lead, in
the ideal case, to less than one candidate over the parameter space
points considered in this search.  

Assuming that the gravitational wave emission mechanism is due to
deviations of the pulsar's shape from perfect axial symmetry, from
Eq.~(\ref{eq:h0}). (Eq.~(2.9) in \cite{hough04}) and
Eq.~(\ref{eq:sensitivity}), we can estimate the nominal
astrophysical reach of the search for the three detectors:
\be \label{eq:range}
d = \frac{16\pi^2GN^{1/4}I_{zz}\epsilon
f_r^2}{8.54c^4}\sqrt{\frac{\Tcoh}{S_n(2f_r)}} \,.
\ee
For a value of $I_{zz} = 10^{45}\,\textrm{g-cm}^2$, $\epsilon =
10^{-5}$, and for typical parameters of the S2 run, this
corresponds to a distance of about $20$-$30$ parsecs.  It should be
kept in mind that this is not a realistic figure for the
astrophysical reach of the search; it does not consider the
mismatch between the template and signal, and it does not use the 
more realistic false alarm rate mentioned above.  Due to these
effects, it turns out that Eq.~(\ref{eq:range}) overestimates the
astrophysical reach by a factor of about $2$-$3$.

\section{The search pipeline}
\label{sec:parameters}

Data from each of the three LIGO interferometers are used to analyze
the same parameter space region.  This section describes the portion
of the parameter space $(f_0,\dot{f},\mathbf{n})$ that we search over,
and the resolution of our grid in this portion of the parameter space.

The total observation time is approximately $\Tobs \approx
5.2\times 10^6$sec corresponding to the S2 science run. We search
for pulsar signals in the frequency range $200$-$400$Hz with a
frequency resolution: $\delta f = {\Tcoh}^{-1} = 5.556\times
10^{-4}\textrm{Hz}$. The resolution $\delta\dot{f}$ in the space
of first spindown parameters is given by the smallest value  of
$\dot{f}$ for which the intrinsic signal frequency does not drift
by more than a single frequency bin during the total observation
time: $\delta \dot f = {\delta f}\cdot \Tobs^{-1}\sim 1.1\times
10^{-10}\textrm{Hz/s}$. We choose the range of values $-\dot
f_\max < \dot f\le 0$, where $\dot f_\max = 1.1\times
10^{-9}\textrm{Hz/s}$. This yields eleven spin-down values.   All
known pulsars (except for a few supernova remnants) have spindown
parameters less than this value. 
This value of $\dot{f}_\max$ is equivalent to looking for pulsars
whose spindown age $\tau = \hat{f}/\dot{f}$ is at least
$1.15\times 10^{4}$ yr.  This also shows that the
approximation to drop higher spindown terms in
Eq.~(\ref{eq:phasemodel}) is reasonable; with a spindown age of
$1.15\times 10^4$ yr as above, we would need a total observation time
of $\sim 10$ yr for the second spindown to cause a frequency drift of
half a frequency bin. 

The resolution $\delta\theta$ in sky positions is frequency dependent,
with the number of templates increasing with frequency and is given by
$\delta\theta = \frac{1}{2}(\delta \theta)_\min$, where
$(\delta\theta)_\min$ is given in Eq.~(4.14) of \cite{hough04}.  This
yields a resolution of about $9.3 \times 10^{-3}$rad at $300$Hz.  This
resolution corresponds to  $ \sim 1.5\times 10^{5}$ sky locations for
the whole sky.

\begin{figure}
  \begin{center}
  \includegraphics[height=10cm]{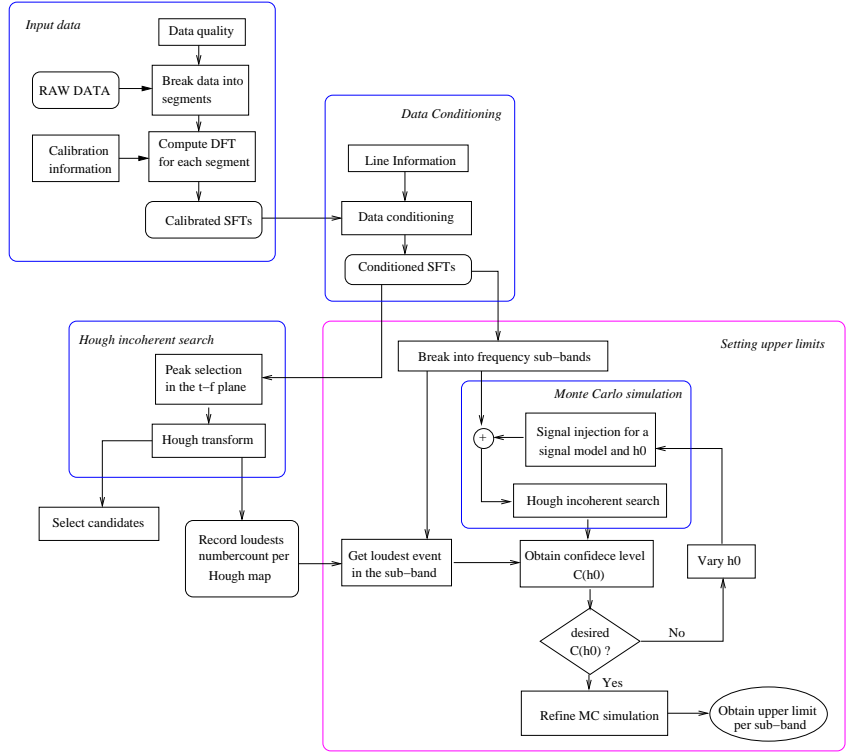}
  \caption{Pipeline for the semi-coherent Hough search for a single interferometer.}
  \label{fig:pipe}
  \end{center}
\end{figure}
The pipeline used to search over this parameter space is described in
figure \ref{fig:pipe}.  The figure is divided into four distinct
blocks.  The top-left block is the preparation of the SFTs: the data
stream is broken up into segments, calibrated, and a discrete Fourier
transform is applied to each segment.  The calibration connects the
error-signal from the interferometer to the actual value of the
strain, and this is calculated in the frequency domain. These SFTs are
passed onto an optional conditioning step.  This is meant to remove
any known spectral disturbances from the SFTs.  In the present paper,
we only present results for which no data conditioning is
applied to the SFTs.  Finally, data from the three interferometers are
analysed separately.  

The rest of the pipeline consists of two conceptually distinct
parts: the actual Hough search and the process of setting upper
limits.  The Hough search has been described earlier; a threshold
is set on the normalized power of each SFT, replacing thereby the
SFTs by a set of peakgrams.  In this paper, we only present partial
results from this Hough search and not the process of setting upper
limits in any detail, except to say that this is the conventional
frequentist upper limit based using Monte-Carlo simulations.  We
set upper limits in each 1Hz frequency band, based on the loudest
event observed in that band. This will be presented elsewhere.

\section{Partial results from the search code}
\label{sec:searchresults}

As described earlier, the first step in this semi-coherent Hough
search is to select frequency bins from the SFTs by setting a
threshold on the normalized power defined in
Eq.~(\ref{eq:normpower}). This requires a reliable estimate of the
power spectral density $S_n$ for each SFT, for which we
employ a running median applied to the periodogram
of each individual SFT. The running median is a robust method
to estimate the noise floor \cite{mohanty02b} which
has the virtue of discarding outliers which appear in a small number
of bins, thereby providing an accurate estimate of the noise floor in
the presence of spectral disturbances and possible signals.
\begin{figure}
  \begin{center}
  \includegraphics[height=8cm]{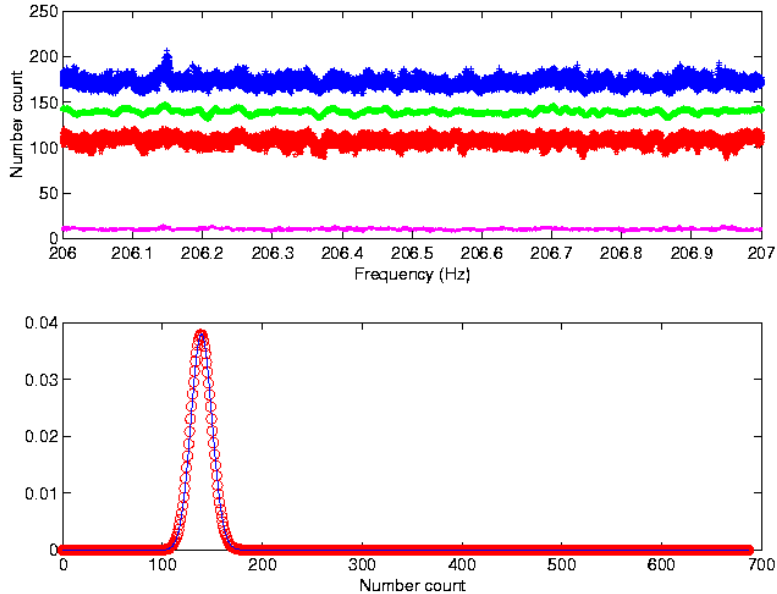}
  \caption{Top: maximum, mean, minimum and standard deviation of the number
  count of all the Hough maps in the frequency band 206-207 Hz. The data
  corresponds to L1 for the entire S2 run using 687 SFTs with a time baseline of
  30 minutes.
  Bottom: the solid line corresponds to the L1 number-count distribution
  obtained in that band, and in red circles the theoretical expected binomial
  distribution for 687 SFTs and a peak selection probability of $20\%$.}
  \label{fig:stats}
  \end{center}
\end{figure}
\begin{figure}
  \begin{center}
  \includegraphics[height=7cm]{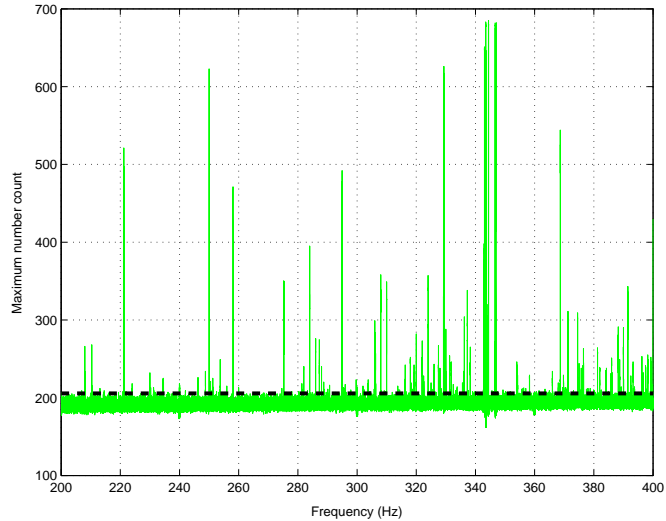}
  \caption{Graph of the L1 maximum  number count per frequency analyzed,
  maximized over all spin-down values and sky locations. The dash-dotted line
  is the corresponding threshold $n_\th$ for a false alarm $\alpha$ of
  $10^{-10}$.  }
  \label{fig:L1max}
  \end{center}
\end{figure}

As an illustrative example, some results of the Hough search in a
1 Hz frequency band are shown in figure
\ref{fig:stats}.  The first panel of this figure shows, for every
frequency bin, the maximum, minimum, mean and standard deviation of
the number counts for all sky-locations and all spindown values.  As
expected, the mean is approximately $Nq = 0.2 \times 687 \approx
137$.  Similarly, as expected, the standard deviation is
$\sqrt{q(1-q)N} \approx 10$.   The second panel of figure
\ref{fig:stats} shows the distribution of number counts in this band
and compares it with the expected binomial distribution in the absence
of any signal.  We find excellent agreement with the expected binomial
distribution, and this is true in all frequency bands which are
relatively free of spectral disturbances.

Figure \ref{fig:L1max} shows the largest number count obtained for
every frequency bin (i.e. the maximum number count over all
sky-locations and spindown values for a given frequency value). 
As this figure shows, several environmental and instrumental noise
sources are present.  The sources of these disturbances are mostly
understood.  They consist of broad $60$Hz power lines, multiples of
$16$Hz due to the data acquisition system, and the violin modes of the
mirror suspensions in a neighborhood of $345$Hz.
The 60Hz lines are rather broad, with a width of about $\pm 0.5$Hz,
while the 16Hz data acquisition lines are confined to a single
frequency bin.  In addition to the above disturbances, we also observe
a large number of multiples of $0.25$Hz.  While these lines are known
to be instrumental, their exact physical origin is yet to be
determined.

\section{Pipeline validation with hardware signal injections}
\label{sec:hardwareinj}

Two artificial pulsar signals were injected for a duration of 12 hours
at the end of the S2 run into all three LIGO interferometers.
These injections were designed to give an end-to-end
validation of the search pipeline starting from as far up the
observing chain as possible.

The two artificial signals were injected at frequencies of
1279.123\,Hz (P1) and 1288.901\,Hz (P2) with
spindown rates of zero and $-10^{-8}$\,Hz\,s$^{-1}$ respectively, and
amplitudes $h_0$ of $2\times 10^{-21}$. The signals were modulated and
Doppler shifted to simulate sources at fixed positions on the sky
with $\psi=0$, $\cos\iota=0$ and $\phi=0$.  P1 was injected at a right
acsension of 5.147 rad and a declination of 0.3767 rad, while P2 had a
right ascension of 2.3457 rad and a declination of 1.2346 rad.  

The resolution in the space of sky positions and frequencies are
the same as in section \ref{sec:parameters}, but the spin-down
resolution depends on the total observation time, and this now
turns out to be $-2.28624\times 10^{-8}$ Hz\,s$^{-1}$ for L1,
$-1.77024\times 10^{-8}$ Hz\,s$^{-1}$ for H1, and $-1.93533\times
10^{-8}$ Hz\,s$^{-1}$ for H2. As before, for each intrinsic
frequency we analyze 10 different spin-down values. The portion of
the sky analyzed has a width of $0.5$ radians $\times$ $0.5$
radians centered around the location of the injected signals.
\begin{figure}
  \begin{center}
  \includegraphics[height=9cm]{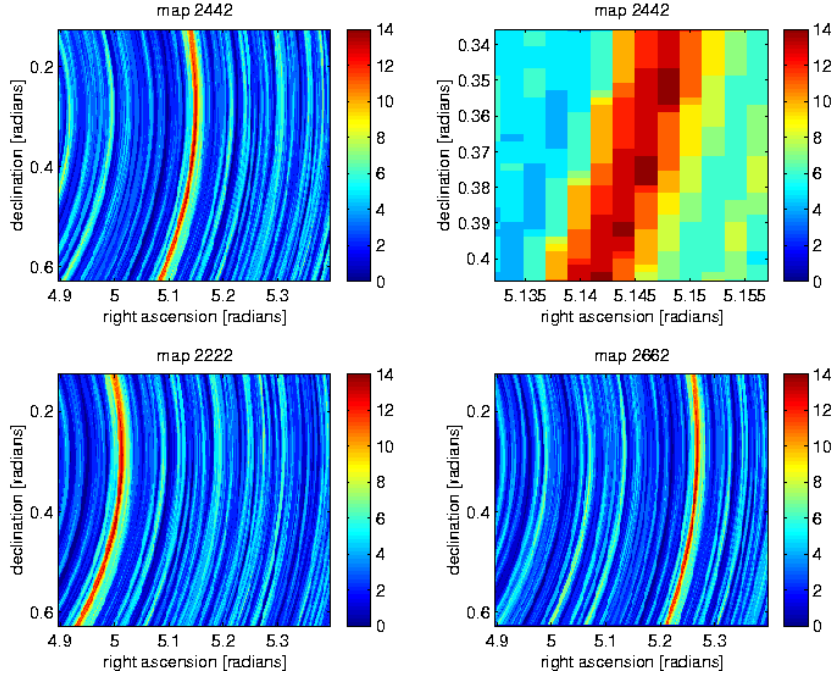}
  \caption{Hough maps for the hardware injected signal P1 in L1. Map
  2442 (top-left) corresponds to 1279.123333 Hz, and contains the
  template which is closest to the signal.  The top-right panel is a
  zoom of this map, showing the signal more clearly.  Maps 2222 and
  2662 (bottom-left and bottom-right) have a larger mismatch in
  frequency; they correspond to 1279.112222 Hz and 1279.134444 Hz
  respectively.  The signal is detected in these maps also, but with a
  mismatched sky-location. P1 was injected at a right
  acsension and declination of 5.147 rad 0.3767 rad respectively.
  This sky-location corresponds roughly to the center of the
  skypatches shown in these figures. 
}\label{fig:injections}
  \end{center}
\end{figure}

Figure \ref{fig:injections} shows some results for pulsar P1 using
L1 data. There were 14 SFTs available in the duration when the
pulsar was injected. The top left and top right panels of figure
\ref{fig:injections} shows Hough maps corresponding to the
frequency and spin-down values nearest to the injected signal.
Although the presence of the signal is clearly visible, 12 hours
of observation time is not enough to identify the location of the
source in the sky. In particular, while the signal is identified
with a high number count in these hough maps, one can still
identify the signal in Hough maps corresponding to different
frequencies and spindowns with high number-counts, but with a
mismatch in the sky location.  This is shown in the bottom left
and bottom right panels of \ref{fig:injections}. Similar results
were found for pulsar P2 and the other detectors, thus providing
an important validation of this search pipeline.

\section{Conclusions}
\label{sec:conc}

In this paper, we have described the idea of the Hough search and the
search pipeline used to analyze data from the second science run of
the LIGO interferometers. We have shown some outputs of the Hough
search pipeline in the frequency range $200$-$400$Hz, over the whole
sky, and the first spindown parameter.  We
have also validated the search pipeline by showing that the search can
detect hardware injected pulsar signals.
Work is in progress to compute astrophysical upper limits using the
search pipeline presented in this paper.

The eventual role of the Hough transform is in a hierarchical scheme
\cite{bc,cgk}.  The Hough transform could be used as a computationally
inexpensive and robust method for quickly scanning large parameter
space volumes and producing significant candidates for a follow-up
search using a more sensitive method.

\section{Acknowledgments}

The authors gratefully acknowledge the support of the United States National Science
Foundation for the construction and operation of the LIGO Laboratory and the
Particle Physics and Astronomy Research Council of the United Kingdom, the Max-Planck-Society
and the State of Niedersachsen/Germany for support of the construction and
operation of the GEO600 detector. The authors also gratefully acknowledge the support
of the research by these agencies and by the Australian Research Council, the
Natural Sciences and Engineering Research Council of Canada, the Council of Scientific
and Industrial Research of India, the Department of Science and Technology of India,
the Spanish Ministerio de Educaci\'on y Ciencia, the John Simon Guggenheim Foundation,
the David and Lucile Packard Foundation, the Research Corporation, and the Alfred P. Sloan Foundation.



\section*{Bibliography}

\begin{thebibliography}{99}

\bibitem{S1-CW}  Abbott B \etal (The LIGO Scientific Collaboration)
  2004 \textit{Phys. Rev. D} {\bf 69} 082004

\bibitem{cw-prl} \dash 2005 \textit{Phys. Rev. Lett.} \textbf{94}
  181103 

\bibitem{astone}  Astone P \etal 2002 \textit{Phys. Rev. D}
  \textbf{65} 042003 

\bibitem{hough04} Krishnan B \etal 2004 \textit{Phys.Rev. D} {\bf 70}
  082001 

\bibitem{jks}  Jaranowski P, Kr\'olak A, and Schutz B F 1998
  \textit{Phys. Rev. D} {\bf 58} 063001

\bibitem{hough1} Hough P V C 1959 \textit{In International Conference
  on High Energy Accelerators and Instrumentation} CERN

\bibitem{hough2} \dash 1962 \textit{U.S. Patent 3,069,654}

\bibitem{ik} Illingworth J and Kittler J 1988 \textit{Computer Vision,
  Graphics, and Image Processing} \textbf{44} 87-116

\bibitem{bc} Brady P R and Creighton T 2000  \textit{Phys. Rev. D}
  {\bf 61} 082001

\bibitem{cgk} Cutler C, Gholami I, and Krishnan B, \texttt{gr-qc/0505082}

\bibitem{ligo-geo} Abbott B \etal 2004 \textit{Nucl. Instrum. Meth. A}
  \textbf{517} 154-179  

\bibitem{mohanty02b} Mohanty S D 2002 \textit{\CQG} \textbf{19} 1513



\end{thebibliography}
\end{document}